\documentclass[aps,prl,twocolumn,superscriptaddress]{revtex4-2}

\usepackage{graphicx}
\usepackage{natbib,hyperref}
\usepackage[english]{babel}
\usepackage{changes}
\usepackage{xcolor}%

\begin{document}

\newcommand{\fig}{Fig.\,}
\newcommand{\GdRuSi}{GdRu\textsubscript{2}Si\textsubscript{2}}
\newcommand{\Q}{$\mathbf{Q}$ }
\newcommand{\QI}{$\mathbf{Q}_1$ }
\newcommand{\QII}{$\mathbf{Q}_2$ }
\newcommand{\Mz}{$\mathbf{M}_\mathrm{z}$ }

\title{SP-STM study of the multi-\Q phases in \GdRuSi}

\author{Jonas~Spethmann}
\email{jspethma@physnet.uni-hamburg.de}
\affiliation{Institute for Nanostructure and Solid State Physics, University of Hamburg, 20355 Hamburg, Germany}
\author{Nguyen Duy Khanh}
\affiliation{Department of Applied Physics and Institute of Engineering Innovation, University of Tokyo, Bunkyo, Tokyo 113-8656, Japan}
\author{Haruto Yoshimochi}
\affiliation{Department of Applied Physics and Institute of Engineering Innovation, University of Tokyo, Bunkyo, Tokyo 113-8656, Japan}
\author{Rina Takagi}
\affiliation{Department of Applied Physics and Institute of Engineering Innovation, University of Tokyo, Bunkyo, Tokyo 113-8656, Japan}
\affiliation{Institute for Solid State Physics, University of Tokyo, Kashiwa 277-8581, Japan}
\affiliation{PRESTO, Japan Science and Technology Agency (JST), Kawaguchi, 332-0012, Japan}
\author{Satoru~Hayami}
\affiliation{Graduate School of Science, Hokkaido University, Sapporo 060-0810, Japan}
\author{Yukitoshi~Motome}
\affiliation{Department of Applied Physics, The University of Tokyo, Tokyo 113-8656, Japan}
\author{Roland~Wiesendanger}
\affiliation{Institute for Nanostructure and Solid State Physics, University of Hamburg, 20355 Hamburg, Germany}
\author{Shinichiro Seki}
\affiliation{Department of Applied Physics and Institute of Engineering Innovation, University of Tokyo, Bunkyo, Tokyo 113-8656, Japan}
\affiliation{PRESTO, Japan Science and Technology Agency (JST), Kawaguchi, 332-0012, Japan}
\author{Kirsten~von~Bergmann}
\email{kirsten.von.bergmann@physik.uni-hamburg.de}
\affiliation{Institute for Nanostructure and Solid State Physics, University of Hamburg, 20355 Hamburg, Germany}
		
\date{\today}

\begin{abstract} 
The two stable surfaces of \GdRuSi\ are studied using spin-polarized scanning tunneling microscopy (SP-STM). Depending on the applied magnetic field different magnetic phases have been found and the presented measurements are in agreement with the respective previously proposed multi-\Q spin textures. In particular the multi-\Q nature of the zero magnetic field state, for which previous experiments could not rule out the coexistence of single-\Q states, can be confirmed by our spin-resolved measurements on the Si-terminated surface. The surfaces of \GdRuSi\ exhibit strong magnetism-induced modulations of the spin-averaged density of states. We find that while the magnetic contribution to the tunnel signal can be clearly identified for the Si-terminated surface this proves to be much more difficult for the Gd-terminated surface. However, the magnetic field dependent spatial modulations on the Gd-terminated surface demonstrate that additional magnetic phase transitions occur for the surface layer compared to those identified for bulk \GdRuSi\ and possible spin textures are presented.
\end{abstract}

\maketitle

\section{Introduction}

Magnetic multi-\Q states such as hexagonal skyrmion lattices can arise at different length-scales, ranging from lattice constants of several ten nanometers~\cite{muhlbauer2009,yuN2010,seki2012} down to one nanometer~\cite{rommingS2013a,vonbergmannNL2015}. Whereas large skyrmions are typically driven by the competition of Heisenberg exchange and Dzyaloshinskii-Moriya (DM) interaction~\cite{nagaosaNN2013}, at the atomic scale higher order interaction between more than two spins become important for the formation of two-dimensionally modulated magnetic states~\cite{kurz_prl_2001,akagi_spin_2010,heinzeNP2011,akagi_hidden_2012,ozawa_zero-field_2017,hayami_effective_2017,hayami_topological_2021}. Atomic-scale multi-\Q states have been found both in transition metal Fe monolayers on non-magnetic metallic substrates~\cite{heinzeNP2011,vonbergmannNL2015}, and in the bulk of the composite material \GdRuSi\ with a rare earth element~\cite{khanh2020}. In the first case the nanoskyrmion lattice is the ground state at zero-magnetic field and no phase transitions have been found up to the available magnetic field of 9T, whereas \GdRuSi\ exhibits a richer phase diagram and the skyrmion lattice arises in an external magnetic field.

\GdRuSi\ crystallizes in the centro-symmetric tetragonal ThCr$_2$Si$_2$ structure type (space group I4$/mmm$)~\cite{hiebl1983, slaski1984} and recent investigations of \GdRuSi\ have revealed three different non-collinear spin structures which occur at different applied magnetic fields~\cite{khanh2020}. Resonant X-ray scattering (RXS) experiments show that each of these magnetic phases contains the same two orthogonal modulation vectors \QI$\approx(0.22, 0, 0)$ and \QII$\approx(0, 0.22, 0)$. Due to the spatially averaging nature of RXS experiments, it is difficult to distinguish if these two modulation vectors are the result of a multi-\Q character of the magnetic phases or if they reflect two rotational domains of a single-\Q state that occur within different segments of the same crystal. However, utilizing Lorentz transmission electron microscopy (TEM), the authors were able to identify a multi-\Q skyrmion lattice structure in the second phase (Phase~II). 

In a follow-up scanning tunneling spectroscopy (STM) study the \GdRuSi\ crystal surface was studied with a spin-averaging STM tip and magnetic field dependent changes in the local density of state (LDOS) were found~\cite{yasui2020}, that were analyzed both in real-space STM images and in their fast Fourier transforms (FFT). The authors argued that due to the interaction between itinerant electrons and localized magnetic moments the magnetic texture has an impact on the LDOS, giving rise to magnetism-related spatial modulations observed in spin-averaging STM. This is similar to what has previously been observed for spin spirals and magnetic skyrmions and was attributed to tunnel anisotropic or non-collinear magnetoresistance~\cite{bode2002,bergmann2012,herve2018,hannekenNN2015,crumNC2015,kubetzka2017}. Because often one cannot identify which mechanism plays the dominant role, in the following we will sum up all these signal variations under the term electronic magnetoresistance (EMR). The STM measurements of Yasui \textit{et al}.\,~\cite{yasui2020} have exploited the EMR to reveal the fourfold symmetry for Phase~II and Phase~III, as expected for the proposed multi-\Q states, however, the question about the multi-\Q nature of Phase~I remained unresolved.

Here, we employ spin-polarized (SP-)STM~\cite{wiesendanger2009,vonbergmannJPCM2014} with a magnetic tip to investigate the \GdRuSi\ surface in the different magnetic phases. In SP-STM the spin-polarized contribution to the tunnel current scales with the angle between tip and sample magnetization, i.e., the tunnel magnetoresistance (TMR) effect in STM geometry is exploited. We provide evidence for the multi-\Q nature of Phase~I, investigate details of the magnetic and electronic contrast components, and discuss differences in the magnetic phase diagram of the Gd-terminated surface layer compared to that of the \GdRuSi\ bulk.

For spin spirals this EMR typically occurs in real space with half of the TMR period~\cite{bode2002,bergmann2012,herve2018,hannekenNN2015}, i.e., $\mathbf{Q}^\mathrm{EMR} = 2 \mathbf{Q}^\mathrm{TMR}$, whereas in two-dimensionally modulated magnetic states as nanoskyrmion lattices it also has contributions at $\mathbf{Q}^\mathrm{EMR} = \mathbf{Q}^\mathrm{TMR}_1 + \mathbf{Q}^\mathrm{TMR}_2$~\cite{heinzeNP2011}. Different EMR patterns have been observed in the same system depending on the used bias voltage~\cite{heinzeNP2011} and in general it has been found that the strength of an EMR contrast is also tip-dependent, suggesting that the orbitals of the tip play a role for the detection of the EMR.

\section{Multi-\Q phases of \GdRuSi}

\begin{figure}[bth]
	\centering
	\includegraphics[width=\columnwidth]{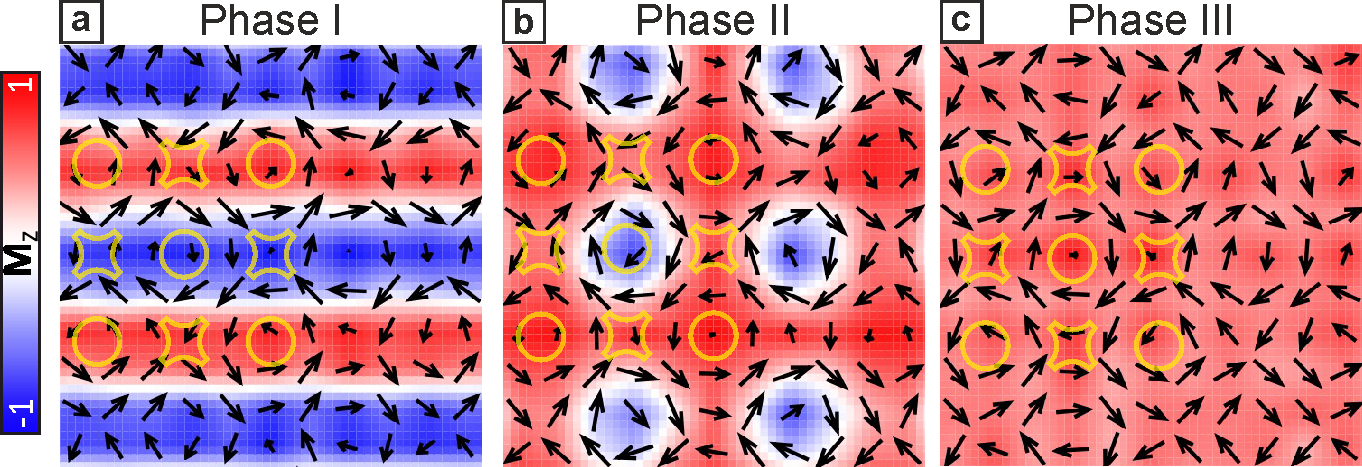}	
	\caption{The magnetic phases of \GdRuSi. (a),(b),(c)~The multi-Q spin structures of bulk \GdRuSi\ according to Khanh \textit{et al}.\,~\cite{khanh2020} and Yasui \textit{et al}.\,~\cite{yasui2020}; note that in contrast to previous sketches the lattice constants of the square magnetic unit cells are 4.5~atomic distances, in agreement with the STM experiments. The displayed spin calculations are based on the model proposed in Ref.\,\cite{hayami_square_2021}. These spin textures can be decomposed into building blocks of vortices (circles) and antivortices (fourfold). The red and blue color indicates the \Mz component. The arrows mark the positions of atomic magnetic moments and their lengths indicate the size of their in-plane components.} 
	\label{fig1}	
\end{figure}

Based on the experimental results of Khanh \textit{et al}.\,~\cite{khanh2020,khanh_zoology_2022} three multi-\Q spin structures for the different phases were proposed, which are sketched in \fig\ref{fig1}(a)-(c). These spin structures can be decomposed into building blocks of vortices and antivortices, as indicated in the images by circles and fourfold objects, respectively. These vortices and antivortices form two intertwined square sublattices, giving rise to square magnetic unit cells with lattice vectors corresponding to \QI and $\mathbf{Q}_2$. Phase~I is characterized by an uniaxial modulation of the \Mz component (colored red for up and blue for downward pointing magnetization), i.e., rows of neighboring vortices and antivortices have the same $\mathbf{M}_\mathrm{z}$, which is opposite to that of the neighboring rows. This results in a spin structure with zero net magnetization and Phase~I is stable up to a magnetic field value of 2.1~T. Phase~II is a square skyrmion lattice, in which all antivortex cores are aligned with the magnetic field direction, whereas vortex cores are unchanged with respect to Phase~I, i.e., half of them point up and the other half down, see \fig\ref{fig1}(b). This results in a net out-of-plane magnetization component and this Phase~II is stable at field values between 2.1~T and 2.6~T. In Phase~III all vortex cores are flipped and point in the direction of the external magnetic field, see \fig\ref{fig1}(c), resulting in an increase of the net magnetization of the spin texture. 

\section{Experimental Results}

The \GdRuSi\ single crystals were grown using the floating zone method and their purity was verified via X-ray diffraction. All experiments were performed in ultra-high vacuum and \GdRuSi\ single crystals were cleaved \textit{in-situ}. \GdRuSi\ consists of a square lattice of Gd\textsuperscript{3+} ions between Ru\textsubscript{2}Si\textsubscript{2}\textsuperscript{3-} layers~\cite{hiebl1983,slaski1984} and it typically cleaves between layers of Gd and Si. Naturally, this will only expose Gd- or Si-terminated surfaces, but never layers of Ru. These layers were identified previously by comparing tunnel spectroscopy data of the different surfaces with DFT calculations of the local density of states (LDOS)~\cite{yasui2020}. In STM measurements both types of surface-terminations can be easily distinguished, because with sufficient image resolution the Si-terminated surface always shows the square atomic lattice ($a = 4.16$~\AA), whereas the Gd-terminated surface typically appears flat.

\begin{figure}[tbh] 
	\centering
	\includegraphics[width=1\columnwidth]{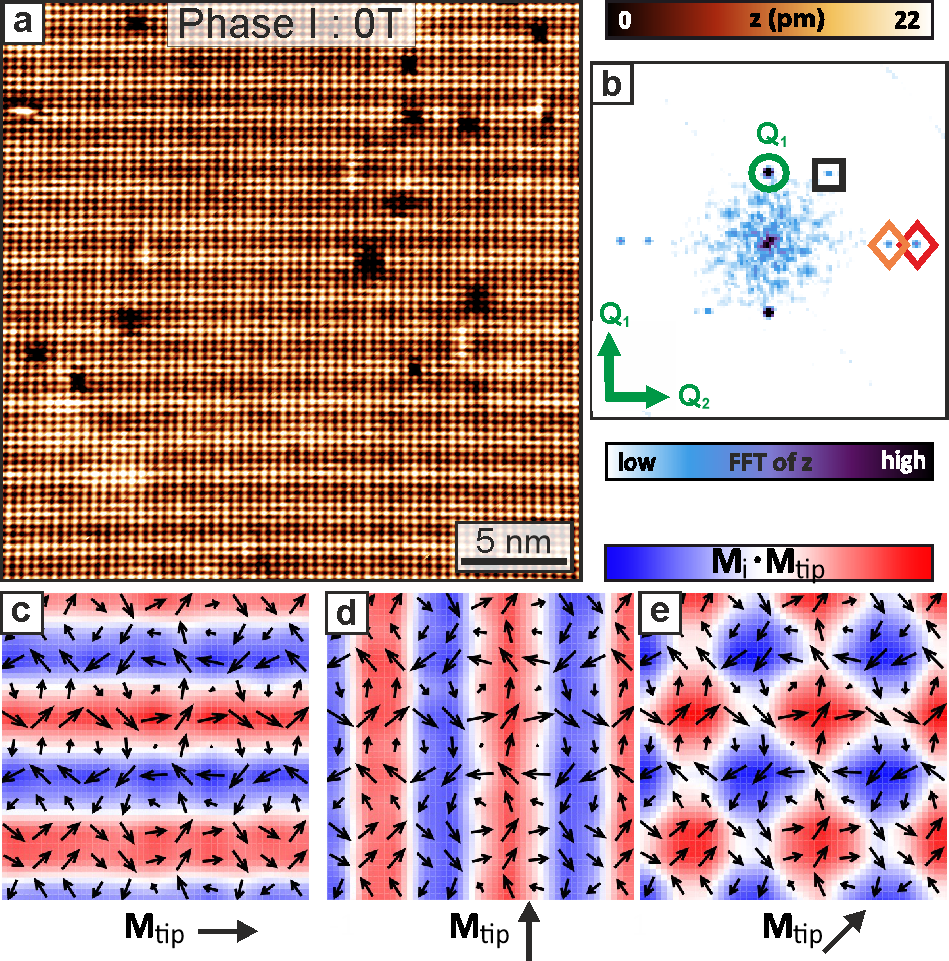}
	\caption{SP-STM measurement of the Si-terminated surface in Phase~I. \textbf(a)~Constant-current SP-STM image with strong spin-polarized contrast measured with in-plane magnetized tip ($U=-20$~mV, $I=100$~pA, $T=8$~K, Fe-coated W tip). \textbf(b)~FFT of (a); the highlighted spots correspond to different modulation vectors: \QI (green circle), \QI+~\QII (black square), 2\QII (orange diamond), and a replica spot at $\mathbf{G}_2 - 2$\QII (red diamond). \textbf(c),(d),(e)~The expected SP-STM contrast of Phase~I for different tip magnetization directions as indicated.}  
	\label{fig2}
\end{figure}

\vspace{3mm}
\noindent\textbf{The Si-terminated surface}

We start by investigating the Si-terminated surface and \fig\ref{fig2}(a) shows an SP-STM constant-current image obtained with an Fe-coated W tip, which is typically magnetized along an arbitrary in-plane direction in zero magnetic field. In this measurement the TMR is exceptionally strong and can easily be seen in the raw data that also contains the atomic corrugation of the Si-terminated surface. It manifests as a sinusoidal contrast modulation, apparent by brighter and darker horizontal stripes. In the FFT of this image, see \fig\ref{fig2}(b), this modulation corresponds to a spot at \QI (green circle). In addition we observe spots at 2\QII (orange diamond) and \QI + \QII (black square). These spots have been previously detected in the study by Yasui~\textit{et al}.\ using a non-magnetic tip and therefore can be at least partially attributed to the EMR. The spot indicated by the red diamond at $\mathbf{G}_2 - 2$\QII is a result of the beating of the atomic periodicity ($\mathbf{G}$) with that of the EMR (see also Supplementary Fig.\,S1).

Figures\,\ref{fig2}(c)-(e) show the expected TMR patterns of Phase~I for different in-plane oriented tips as indicated. Since \QI and \QII are orthogonal, a tip that is fully aligned with the modulation direction of \QI cannot detect a spin modulation along \QII and vice versa. Naturally, if the tip magnetization axis lies somewhere between both directions, both modulation vectors can be detected as in (e). The experimentally observed TMR pattern for the in-plane magnetized tip used in the measurement of \fig\ref{fig2}(a) agrees nicely with the proposed spin texture and a tip magnetization as sketched in (c). However, while this measurement clearly demonstrates the spin spiral nature of the magnetic state it does not provide evidence for a multi-\Q state. Only the observation of a two-dimensionally modulated pattern as displayed in (e) can prove the multi-\Q nature of Phase~I.

\begin{figure}[tbh] 
	\centering
	\includegraphics[width=0.9\columnwidth]{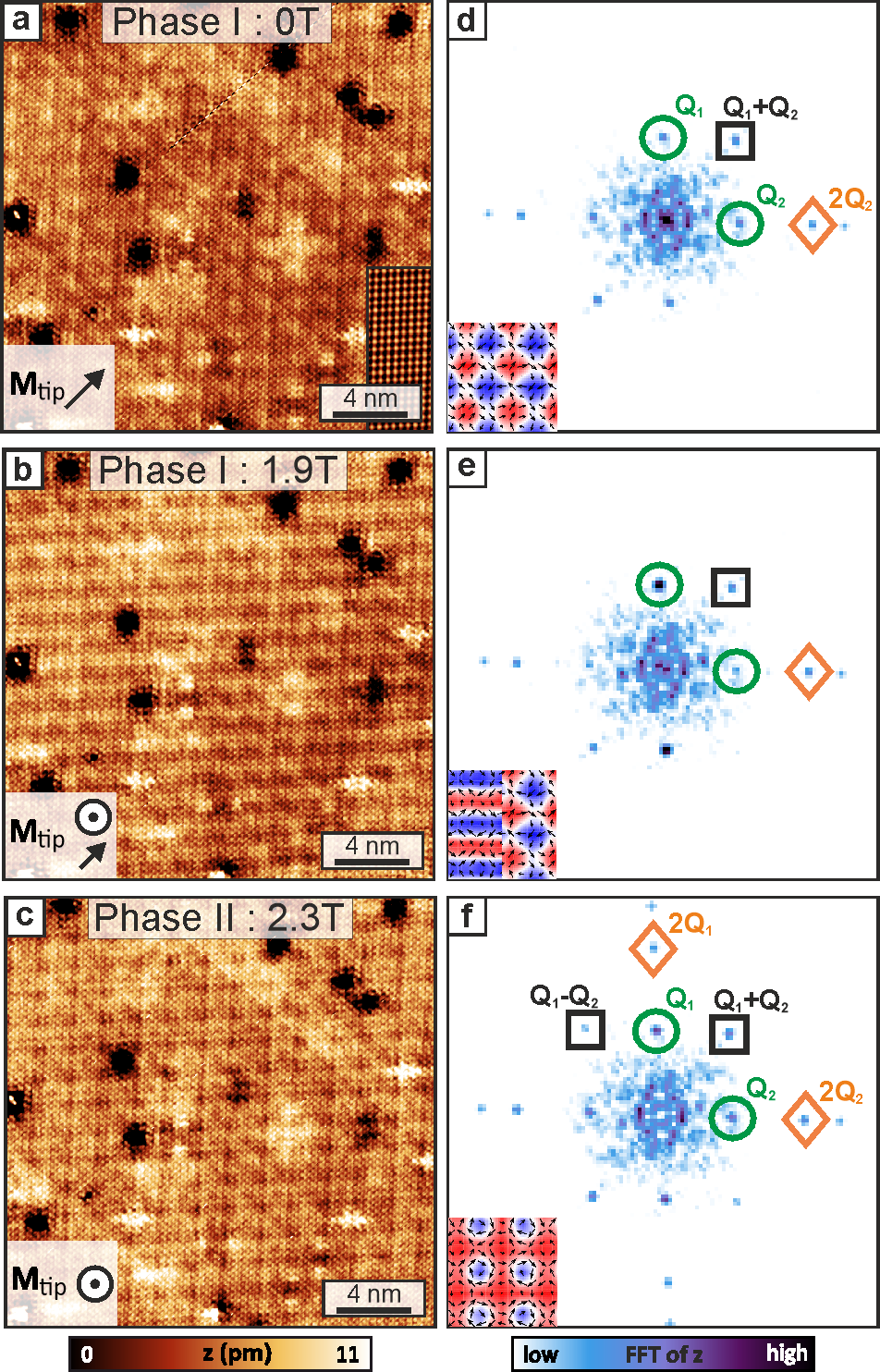}
	\caption{Magnetic field dependent SP-STM measurement on the Si-terminated surface. \textbf(a) Constant-current SP-STM image in zero magnetic field, for which the atomic corrugation (see bottom right inset) has been removed, see Supplementary Fig.\,S2. \textbf(b)~Same at 1.9~T. \textbf(c)~Same at 2.3~T, i.e., in Phase~II. The insets show the tentative tip magnetization direction, which is in-plane at $B=0$~T and is rotated towards the out-of-plane for increasing magnetic fields. \textbf(d),(e),(f)~Corresponding FFTs of (a),(b),(c). (Measurement parameters $U=-20$~mV, $I=100$~pA, $T=8$~K, Fe-coated W tip).} 
	\label{fig3}
\end{figure}

A measurement of Phase~I with a different Fe-coated W tip is shown in \fig\ref{fig3}(a); for this measurement the TMR contribution to the signal is much smaller and the atomic corrugation (see bottom right) is removed for better visibility (see Supplementary Fig.\,S2). The pattern observed in this measurement is in agreement with the TMR displayed in \fig\ref{fig2}(e), providing evidence for the multi-\Q nature of the magnetic texture. Indeed the tip magnetization appears to be along the diagonal of the magnetic unit cell, which is also evident from the equal intensity of the magnetic spots \QI and \QII in the FFT, see green circles in \fig\ref{fig3}(d). The simultaneous observation of \QI and \QII in our SP-STM measurements clearly demonstrates the previously debated multi-\Q nature of Phase~I. 

At an applied out-of-plane magnetic field of $1.9$~T, see \fig\ref{fig3}(b), the sample is still in Phase~I, however, the tip magnetization direction is expected to be partly aligned with the applied field. This means that while in \fig\ref{fig3}(a) the tip is sensitive to the in-plane components of Phase~I, in \fig\ref{fig3}(b) we are also picking up some of the out-of-plane components. Indeed, looking at the FFT in \fig\ref{fig3}(e) we see that the intensity of the magnetic spots \QI and \QII is not equal anymore, suggesting that the TMR pattern can be best described by a combination of out-of-plane and in-plane magnetization components (as in \fig\ref{fig1}(a) and \fig\ref{fig2}(e), respectively), i.e., originating from a canted tip magnetization. The spots marked with the orange diamond remain unchanged, which supports their interpretation as EMR contrast contributions.

At $B=2.3$~T, see \fig\ref{fig3}(c), the \GdRuSi\ crystal is in Phase~II and the tip magnetization is expected to be mostly aligned with the field, i.e., it is sensitive to the out-of-plane sample magnetization components. The TMR contrast shows a lattice with fourfold symmetry, similar to the simulated SP-STM contrast for Phase~II, see \fig\ref{fig1}(b). The intensity of the magnetic spots \QI and \QII is again equal, in accordance with the fourfold symmetry of the \Mz components of Phase~II. Also the EMR related spots at larger \Q (see orange and black markers) now reflect the fourfold symmetry of the magnetic state. In contrast to the mere changes of the spot intensity, the phase transition from Phase~I to Phase~II is reflected by an additional spot at \QI$-$ \QII and a related change to fourfold symmetry.

\vspace{3mm}\noindent\textbf{Comparison of Si- and Gd-terminated surfaces}

\begin{figure}[tb] 
	\centering
	\includegraphics[width=0.9\columnwidth]{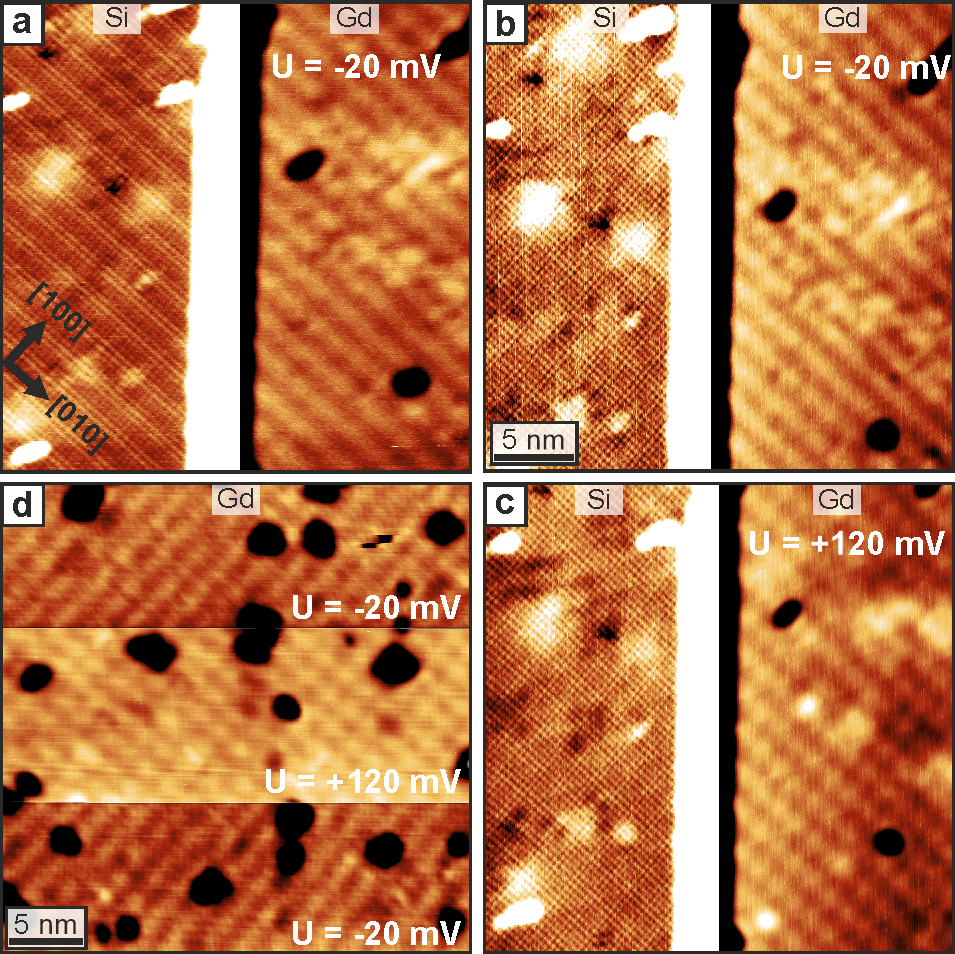}
	\caption{Si- and Gd-terminated surfaces. \textbf(a)~Constant-current SP-STM measurement of a sample area with a Si-terminated terrace adjacent to a Gd-terminated terrace; note that the color contrasts were adjusted separately. \textbf(b)~Same sample area with a different micro-tip. \textbf(c)~Same as (b) but at different bias voltage as indicated. \textbf(d)~Constant current image of a Gd-terminated sample area; during the image the bias voltage was changed as indicated. (Measurement parameters: $U$ as indicated, $I=100$~pA, $T=8$~K, Fe-coated W tip).}  
	\label{fig4}
\end{figure}

In the previous STM study by Yasui~\textit{et al}.\, it was reported that for the Gd-terminated surface additional phase transitions occur~\cite{yasui2020}, which is why we also investigated the Gd-terminated surface using a spin-polarized tip. Figure\,\ref{fig4}(a) shows a sample area where a Gd- and a Si-terminated terrace are separated by a step edge. The height contrast was adjusted individually for the two terraces. In contrast to the Si-terminated terrace, the Gd-terminated surface typically shows no atomic corrugation. Both terraces show a strong contrast modulation that can be described by {$\mathbf{Q}_1$ }, i.e., it is comparable to the strong TMR signal observed for the Si-terminated sample area of \fig\ref{fig2}(a). Figure\,\ref{fig4}(b) shows the same sample area after a gentle tip-surface contact. With this slightly changed tip the magnetic contribution to the signal nearly vanishes for the Si-terminated surface, however, the pattern on the Gd-surface remains equally strong as in the previous image. This suggests that the large contrast on the Gd-terminated surface is not of magnetic origin. Indeed, if the same area is measured with a different bias voltage, see Fig.~4(c), the pattern on the Gd changes, whereas still basically no pattern apart from the atomic lattice is seen on the Si-terminated surface. This bias dependence of the Gd-terminated surface is demonstrated again in \fig\ref{fig4}(d), where the bias voltage was changed during imaging as indicated. We conclude that the observed patterns on the Gd-terminated surface are dominated by bias-dependent EMR patterns. In contrast to the Si-terminated surface the EMR on the Gd appears much more washed out and most importantly it occurs with \QI and \QII and therefore it cannot be distinguished from a TMR pattern.

\vspace{3mm}\noindent\textbf{The Gd-terminated surface}

\begin{figure*}[tb] 
	\centering
	\includegraphics[width=0.8\textwidth]{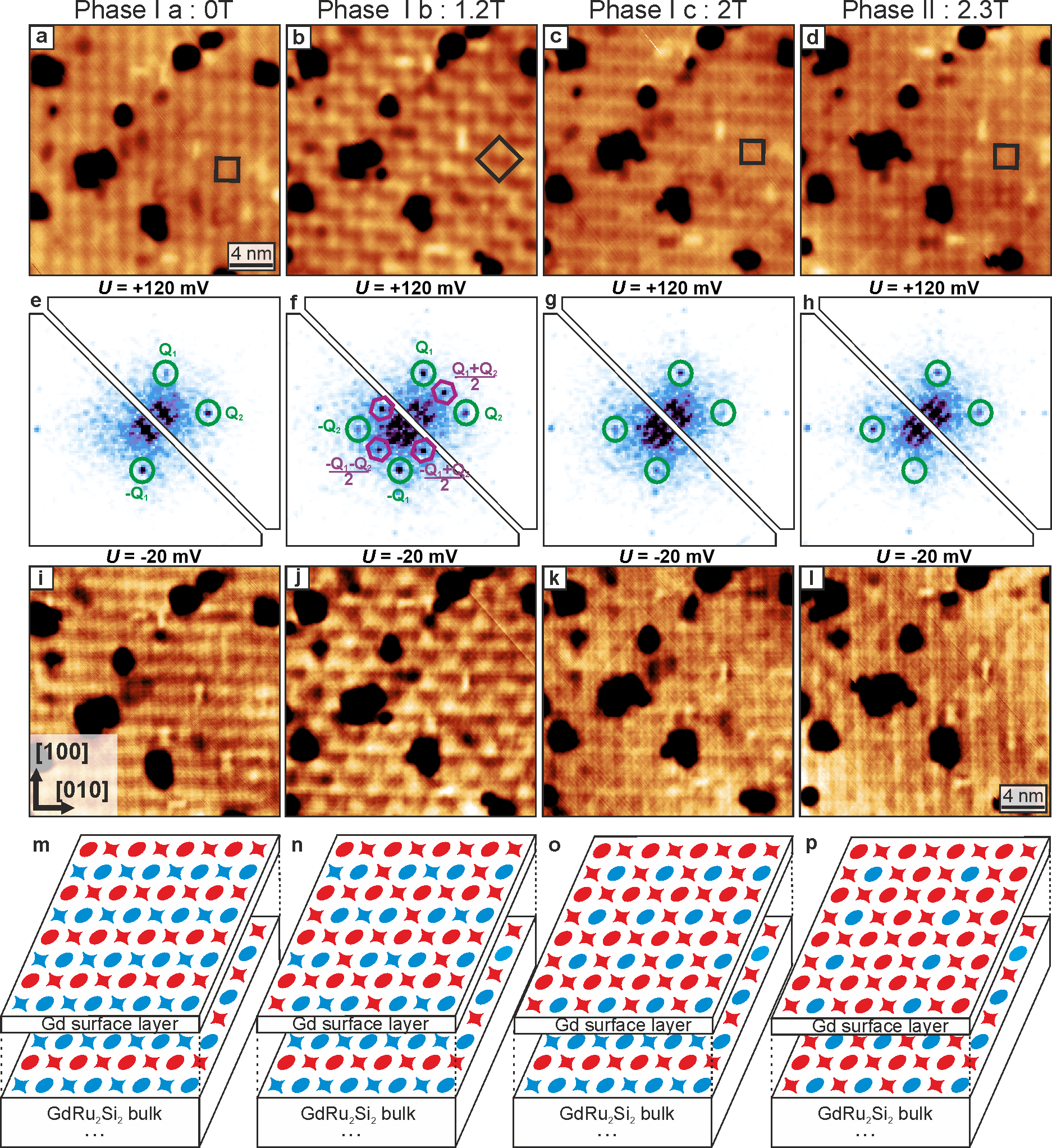}
	\caption{Magnetic field dependent SP-STM measurements of the Gd-terminated surface. \textbf(a),(b),(c),(d)~Constant-current images of the same Gd-terminated surface area in different external magnetic fields; the black squares indicate the unit cells of the observed contrast patterns. \textbf(e),(f),(g),(h)~FFTs of the data shown in (a)-(d) and (i)-(l). \textbf(i),(j),(k),(l)~Constant-current images measured simultaneously to those in (a)-(d). (Measurement parameters: $U$ and $B$ as indicated, $I=100$~pA, $T=8$~K, Fe-coated W tip). \textbf(m),(n),(o),(p) Sketches of tentative magnetic field dependent phases of the Gd-terminated surface together with the respective bulk \GdRuSi\ phases; circles indicate vortices and fourfold objects antivortices, red and blue color indicate the \Mz at the core.}  
	\label{fig5}
\end{figure*}

Figure\,\ref{fig5} shows a measurement series on a Gd-terminated surface area and the first three measurements are at different magnetic field values within Phase~I of bulk \GdRuSi: for the last measurement at 2.3~T the bulk \GdRuSi\ exhibits Phase~II. Figure\,\ref{fig5}(a)-(d) and \fig\ref{fig5}(i)-(j) were measured simultaneously with bias voltages of $+120$~mV in the forward and $-20$~mV in the backward scan, respectively. Each of these images shows a different pattern. Figures\,\ref{fig5}(e)-(h) display the corresponding FFTs and all of them exhibit \QI and \QII spots, except for the FFT of \fig\ref{fig5}(i), which shows the stripe pattern characteristic for the zero-field state at $-20$~mV (see \fig\ref{fig4}).  

The different real space patterns for the three different values of applied field within Phase~I are accompanied by the appearance of additional spots in the FFT at $B = 1.2$~T at $(\mathbf{Q}_1 + \mathbf{Q}_2 ) /2$, indicated with purple hexagons. This is linked to a real space unit cell with twice the area and a $45^\circ$ rotation compared to those at other magnetic fields. Independent of the tip magnetization or the ratio of TMR and EMR contributions, none of the previously discussed spin structures is compatible with the characteristics observed at $1.2$~T. This implies that indeed within Phase~I of the \GdRuSi\ bulk there are at least two additional phase transitions in the Gd-terminated surface layer, one between 0~T and 1.2~T and another one between 1.2~T and 2~T; we call the respective phases Ia, Ib, and Ic.

To understand this discrepancy of the number of phase transitions in the bulk and at the Gd-terminated surface we take a closer look at the environment of the Gd atoms in the two cases, as the Gd atoms provide the magnetic moments that form the spin texture. In the \GdRuSi\ bulk the Gd atoms are positioned between layers of $\textrm{Ru}_2\textrm{Si}_2$. This is not the case for the Gd-terminated surface, where the Gd atoms have a $\textrm{Ru}_2\textrm{Si}_2$ layer only on one side, but are exposed to vacuum on the other side. In consequence the Gd orbitals are expected to hybridize differently with the asymmetric surrounding, which can lead to different magnetic interaction energies and facilitate the stabilization of different magnetic ground states compared to the bulk. In particular, due to the reduced number of neighbors the effective magnetic interaction at the surface is expected to be smaller compared to that in the bulk. In addition, the effective spin-charge coupling is also expected to be reduced by the deformation of the lattice structure and orbital states at the surface. These would also result in the smaller critical magnetic field for phase transitions. While the spin texture of the Gd-terminated surface layer might be different to that of the \GdRuSi\ bulk it does not mean that they are independent of each other, because an exchange coupling between surface and subsurface layers is expected to link the magnetic properties of the two, distort the spin texture in the Gd-terminated surface layer or impose the magnetic periodicity of the bulk onto the surface.

To illustrate this scenario we show in \fig\ref{fig5}(m)-(p) tentative schematic spin configurations for the additional phases in the Gd-terminated surface. Below the Gd surface layer the established spin configurations inside the \GdRuSi\ bulk are sketched in the same fashion for the corresponding fields, with the basic structure of two intertwined lattices of vortices and antivortices with opposite \Mz components in their center (blue and red). For Phase~Ia at $B=0$~T, \fig\ref{fig5}(m), we tentatively keep the bulk spin structure also for the surface Gd layer. For Phase~Ib at 1.2~T, Fig.~5(n), we propose that half of the vortices that were antiparallel to the applied field have switched their \Mz components (from blue to red) to gain Zeeman energy. The resulting larger magnetic unit cell is compatible with the experimental results of Phase~Ib. A further increase of external magnetic field is expected to align further building blocks, and in the sketch of Phase~Ic in \fig\ref{fig5}(o) the Gd surface layer has the same spin texture as the Phase~II bulk phase, however, the bulk is still in Phase~I. This field-dependent scenario of successive switching of building blocks of the surface Gd layer allows for one more step, as sketched in \fig\ref{fig5}(p), which would explain the absence of fourfold symmetry in (i), even though the \GdRuSi\ bulk is now in Phase~II. However, direct evidence of the spin texture of the \GdRuSi\ surface layer cannot be obtained due to the strong EMR contrast which disguises the TMR contribution. We speculate that 4$f$ elements like Gd can in general have much larger impact on the electronic states that are involved in the tunnel process as compared to the 3$d$ elements like Fe studied previously~\cite{bergmann2012} in the context of EMR signatures. 

\section{Conclusion}

We have used SP-STM to study {GdRu\textsubscript{2}Si\textsubscript{2}}, which belongs to a material class of magnets that has not been studied before with this technique. The presented SP-STM measurements on the Si-terminated surface show both TMR and EMR contrasts and allow to prove the multi-\Q nature of Phase~I. All other SP-STM experiments confirm the previously proposed spin textures for Phase~II and Phase~III. For the Gd-terminated surface a disentanglement of TMR and EMR has not been achieved due to the unusually large bias-dependent EMR contribution which masks the TMR in this system due to the same periodicities. However, the measurements have shown two additional phase transitions for the Gd surface layer within a magnetic field regime where the \GdRuSi\ bulk is in Phase~I, and tentative spin textures have been proposed.

\vspace{3mm}

\begin{acknowledgments}
We acknowledge fruitful discussions with Tetsuo Hanaguri, Yuuki Yasui, and Christopher J.\ Butler from RIKEN, Japan, and technical support from André Kubetzka from the University of Hamburg, Germany. K.v.B.\ acknowledges financial support from the Deutsche Forschungsgemeinschaft (DFG, German Research Foundation) via Projects No.~402843438 and No.~418425860. This work was partly supported by Grants-In-Aid for Scientific Research (grant nos 18H03685, 20H00349, 21H04440, 21H04990, 21K13876, 21K18595, 22H04965, 22H04468, 22KJ1061, 23K13069) from JSPS, PRESTO (grant nos JPMJPR18L5, JPMJPR20B4, JPMJPR20L8) and CREST (grant no. JPMJCR23O4) from JST, Katsu Research Encouragement Award and UTEC-UTokyo FSI Research Grant Program of the University of Tokyo, Asahi Glass Foundation and Murata Science Foundation. S.H.\ was supported by JSPS KAKENHI Grants Numbers JP21H01037, JP23H04869, and by JST PRESTO (JPMJPR20L8).
\end{acknowledgments}

\end{document}